\documentclass{article}

\usepackage{PRIMEarxiv}

\usepackage[utf8]{inputenc} % allow utf-8 input
\usepackage[T1]{fontenc}    % use 8-bit T1 fonts
\usepackage{hyperref}       % hyperlinks
\usepackage{url}            % simple URL typesetting
\usepackage{booktabs}       % professional-quality tables
\usepackage{amsfonts}       % blackboard math symbols
\usepackage{nicefrac}       % compact symbols for 1/2, etc.
\usepackage{microtype}      % microtypography
\usepackage{lipsum}
\usepackage{fancyhdr}       % header
\usepackage{graphicx}       % graphics
\graphicspath{{media/}}     % organize your images and other figures under media/ folder

\title{LLM Harmony: Multi-Agent Communication for Problem Solving
}

\author{
  Sumedh Rasal \\
  Georgia Institute of Technology \\
  Chicago, IL\\
  \texttt{srasal3@gatech.edu} \\
}

\begin{document}
\maketitle

\begin{abstract}
% Large Language Models have revolutionized natural language processing, yet challenges persist with occasional inaccuracies and incomplete responses. Multi-agent systems, where agents based on large language models collaboratively tackle tasks, have shown promise. However, to enhance collaborative problem-solving, we propose an innovative approach employing multiple agents. Each agent defines a specific role and provides transparent reasoning for its contributions. In this paper, we delve into the shortcomings of single-agent large language models, showcase the effectiveness of existing multi-agent systems, and introduce a novel framework that promises to outperform by redefining roles and explicating the reasoning behind each agent's contribution. Our approach not only addresses the limitations of current systems but also establishes a clearer and more efficient path to solving complex problems through collaborative reasoning

Large Language Models (LLMs) have revolutionized Natural Language Processing but exhibit limitations, particularly in autonomously addressing novel challenges such as reasoning and problem-solving. Traditional techniques like chain-of-thought prompting necessitate explicit human guidance. This paper introduces a novel multi-agent communication framework, inspired by the CAMEL model, to enhance LLMs' autonomous problem-solving capabilities. The framework employs multiple LLM agents, each with a distinct persona, engaged in role-playing communication, offering a nuanced and adaptable approach to diverse problem scenarios. Extensive experimentation demonstrates the framework's superior performance and adaptability, providing valuable insights into the collaborative potential of multiple agents in overcoming the limitations of individual models.

% \lipsum[1]
\end{abstract}

% keywords can be removed
% \keywords{First keyword \and Second keyword \and More}

\section{Introduction}

The rise of Large Language Models (LLMs) \cite{brown2020language} \cite{touvron2023llama} \cite{thoppilan2022lamda} \cite{patil2023gorilla} \cite{openai2023gpt} \cite{bubeck2023sparks} has undeniably revolutionized the software industry, particularly in the realm of Natural Language Processing (NLP) \cite{devlin2018bert}. These models have demonstrated a remarkable ability not only to generate text but also to grasp the underlying structures of written language, extending their utility to tasks ranging from text generation to code understanding \cite{chen2021evaluating} \cite{schick2023toolformer}. However, despite their prowess, it has become evident that LLMs, while adept at handling familiar topics, often falter when confronted with novel challenges.

One notable limitation lies in their tendency to hallucinate information when presented with unfamiliar subjects \cite{azamfirei2023large}. Moreover, while these models exhibit commendable proficiency in addressing technical coding challenges across multiple programming languages, they struggle with fundamental reasoning questions. Addressing such limitations necessitates an approach that goes beyond conventional methodologies \cite{rajani2019explain} \cite{ling2017program} \cite{cobbe2021training} \cite{chiang2018semantically} \cite{amini2019mathqa} \cite{chen2019neural} \cite{roy2016solving} \cite{lu2023chameleon} \cite{yao2022react}.

In response to this challenge, our paper introduces a novel strategy aimed at enhancing LLM performance on novel problems. Drawing inspiration from the effectiveness of chain-of-thought prompting \cite{wei2022chain} in breaking down complex problems, we seek to leverage the synergy of multiple LLM agents working collaboratively \cite{cohen2023lm} \cite{du2023improving}. Each agent is endowed with a distinct persona, a concept inspired by the CAMEL framework, and engages in role-playing communication methods.

Unlike traditional approaches that might be limited to binary agent personas, such as "user" and "assistant," or simplistic positive and negative agent roles, our proposed framework adopts a nuanced and flexible strategy. The intricacies of various novel scenarios demand a more sophisticated approach. Our design incorporates industry best practices to guide the creation of diverse agent personas, ensuring adaptability to a multitude of problem-solving contexts.

The paper not only addresses the limitations of existing LLM models but also explores the potential of harnessing the collective intelligence of multiple agents to tackle a broader range of challenges. We aim to demonstrate that the proposed framework not only outperforms traditional methodologies but also provides a foundation for autonomous problem-solving, minimizing the need for explicit human guidance.

In the subsequent sections, we delve into the existing techniques that have been developed to address similar challenges, highlighting their strengths and limitations \cite{dafoe2021cooperative}. Following this, we present our multi-agent communication design, emphasizing its improvements over current methodologies \cite{saunders2022self}. The extendability of our approach to various LLM models is a key focus, underlining its potential as a versatile and reusable solution. Through rigorous experimentation and analysis, we aim to showcase the effectiveness and adaptability of our proposed framework in enhancing the quality of LLM outputs across a spectrum of tasks.

This is what we propose:
\begin{itemize}
\item An innovative framework employing multiple large language model workers/agents, each characterized by a unique persona and guided by a chain-of-thought prompt.
\item The collaborative effort of all agents is directed toward devising solutions for novel problems.
\item The framework's versatility allows the incorporation of any persona and chain-of-thought prompt, aligning with the specific problem to be addressed.
\item The framework is available. \url{https://github.com/sumedhrasal/simulation}. It is built on top of CAMEL's and ChatDev's framework.
\end{itemize}

\section{Methodology}
In the realm of collaborative problem-solving, particularly evident in scenarios like software development, the complexities arise from coordinating efforts among diverse individuals. Software development teams, for instance, often undergo iterative cycles before converging on a viable solution, and even then, compromises may be accepted as the best available option \cite{callison2009fast} \cite{liu2023your}. Our proposed strategy addresses the need for comprehensive validation of Large Language Models (LLMs), not just in code evaluation but also in complex reasoning and arithmetic challenges \cite{celikyilmaz2020evaluation} \cite{rouge2004package} \cite{kondrak2005n} \cite{novikova2017we} \cite{wei2022emergent}. Focused on breaking down problem statements and leveraging distinct personas \cite{li2023camel}, the framework provides agents with tailored chain-of-thought prompts \cite{wei2022chain}, enabling multi-agent models to tackle novel problems collaboratively.

Understanding personas is crucial; in a software development organization, roles such as CEO, VP of Engineering, developers, testers, and product managers contribute unique perspectives to problem-solving \cite{qian2023communicative}. The interplay of these personas allows agents to validate each other's responses effectively \cite{woolley2010evidence} \cite{luppi2022synergistic} \cite{wu2023large}. The concept of chain-of-thought mirrors human problem-solving strategies: decomposing a problem, identifying solutions to sub-problems, and constructing a comprehensive answer. The chain-of-thought prompt extends this strategy, offering the LLM an input, a problem-solving approach, and the desired output. The fusion of personas \cite{li2023camel} and chain-of-thought \cite{wei2022chain} prompts capitalizes on mimicking human analytical and execution strategies \cite{ouyang2022training} \cite{sanh2021multitask} \cite{liu2023gpteval} \cite{chiang2023can} \cite{gao2023human} \cite{shen2023large} \cite{zheng2023judging} \cite{sap2019socialiqa} \cite{talmor2018commonsenseqa} \cite{sclar2023minding} \cite{sap2022neural}.

While some might question the necessity of this approach over retraining the model for novel problems, the practicality lies in the belief that emulating human-like analysis and execution strategies can create autonomous agents capable of diverse tasks without extensive model retraining \cite{karpinska2021perils} \cite{van2019best}. Additionally, the high cost and challenges associated with retraining LLMs \cite{rasal2023beyond}, particularly for problems with limited training data, support the appeal of our strategy.

In the realm of multi-agent communication \cite{liang2023encouraging} \cite{du2023improving}, adherence to assigned personas and chain-of-thought prompts is foundational to mitigate agent hallucination and improving cooperation \cite{dafoe2021cooperative} \cite{dafoe2020open}. The framework's versatility extends beyond problem-solving to applications in multi-player games \cite{susskind1985scorable} \cite{susskind2000using} \cite{lazaridou2020multiagent} \cite{graesser2020emergent} \cite{lee2018emergent} \cite{hendrycks2021would} \cite{zhou2023cast} \cite{pan2023rewards} \cite{kramar2022negotiation}, accommodating both team-oriented and individual goal-driven scenarios, and offering flexibility in achieving optimal outcomes.

% \label{sec:headings}

% \lipsum[4] See Section \ref{sec:headings}.

% \subsection{Headings: second level}
% \lipsum[5]
% \begin{equation}
% \xi _{ij}(t)=P(x_{t}=i,x_{t+1}=j|y,v,w;\theta)= {\frac {\alpha _{i}(t)a^{w_t}_{ij}\beta _{j}(t+1)b^{v_{t+1}}_{j}(y_{t+1})}{\sum _{i=1}^{N} \sum _{j=1}^{N} \alpha _{i}(t)a^{w_t}_{ij}\beta _{j}(t+1)b^{v_{t+1}}_{j}(y_{t+1})}}
% \end{equation}

% \paragraph
% \lipsum[7]

% \section{Examples of citations, figures, tables, references}
% \label{sec:others}
% \lipsum[8] \cite{brown2020language, rasal2023beyond} and see \cite{bubeck2023sparks}.

% The documentation for \verb+natbib+ may be found at
% \begin{center}
%   \url{http://mirrors.ctan.org/macros/latex/contrib/natbib/natnotes.pdf}
% \end{center}
% Of note is the command \verb+\citet+, which produces citations
% appropriate for use in inline text.  For example,
% \begin{verbatim}
%    \citet{hasselmo} investigated\dots
% \end{verbatim}
% produces
% \begin{quote}
%   Hasselmo, et al.\ (1995) investigated\dots
% \end{quote}

% \begin{center}
%   \url{https://www.ctan.org/pkg/booktabs}
% \end{center}

\section{Experiments}
In our experimentation, we adopt a two-agent strategy comprising an expert agent and an evaluator agent, both instances of OpenAI's "gpt3.5-turbo." Each agent is configured with specific parameters: temperature set to 0.0, representing the trade-off between coherence and creativity, and a conversation limit of 5, indicating the maximum iterations allowed for inter-agent conversation. The evaluator agent assesses the responses generated by the expert agent, guiding it to rectify inaccuracies. Leveraging OpenAI's LLM object, our framework, integrated with CAMEL's architecture, facilitates seamless communication among multiple LLM agents, aiming to establish a collaborative problem-solving capability for advanced LLM models.

\subsection{Arithmetic Reasoning}
In the arithmetic reasoning segment, we address challenges that assess the arithmetic proficiency of LLMs, an area where traditional large language models have historically faced limitations. The experiment employs the GSM8K benchmark and the SVAMP data set, specifically designed for math word problems. We evaluate the performance of three LLMs: a standalone GPT-3 model, a multi-agent GPT-3 model, and a multi-agent GPT-3 model integrated into our framework. The objective is to gauge the effectiveness of our collaborative approach in enhancing arithmetic reasoning capabilities.

\begin{quote}
    "GSM8K consists of 8.5K high-quality grade school math problems created by human problem writers. We segmented these into 7.5K training problems and 1K test problems. These problems take between 2 and 8 steps to solve, and solutions primarily involve performing a sequence of elementary calculations using basic arithmetic operations (+ - / *) to reach the final answer. A bright middle school student should be able to solve every problem."\cite{cobbe2021gsm8k}
\end{quote}

\begin{quote}
    "We first show that existing models achieve reasonably high accuracies on these datasets even after removing the "question" part of the MWP at test time. We further show that a simple model without any word-order information can also solve a majority of MWPs in these datasets. Our experiments indicate that existing models rely on shallow heuristics in benchmark MWP datasets for achieving high performance. Our experiments render the benchmark datasets unreliable to measure model performance. To enable more robust evaluation of automatic MWP solvers, we created a challenge set called "SVAMP"."\cite{patel2021nlp}
\end{quote}

\subsubsection{Example: 1}\label{Example1.1}
\begin{verbatim}
| Role Name | Student |
| Chain of Thought Prompt | You are the student. 
You will be given a math word problem, your job is to solve this problem. 
Use this template to solve math word problems. 
Input: Roger has 5 tennis balls. He buys 2 more cans of tennis balls. 
Each can has 3 tennis balls. How many tennis balls does he have now? 
Explanation: Roger started with 5 balls. 
2 cans of 3 tennis balls each is 6 tennis balls. 5 + 6 = 11.  
Answer: 11 |
\end{verbatim}

\begin{verbatim}
| Role Name | Teacher |
| Chain of Thought Prompt | You are the teacher. 
You will supply the math word problem to the student agent. 
Once you receive the student agent's answer, compare it against the final answer. 
The correct answer is 20. 
Let the student agent know if his answer is correct or not. |
\end{verbatim}

\begin{verbatim}
Teacher: **[Start Chat]**
Roger has 10 tennis balls. 
He buys 2 more cans of tennis balls. 
Each can has 5 tennis balls. 
How many tennis balls does he have now?

Student response: 
Input: Roger has 10 tennis balls. 
He buys 2 more cans of tennis balls. 
Each can has 5 tennis balls. 
How many tennis balls does he have now?
Explanation: Roger started with 10 balls. 
2 cans of 5 tennis balls each is 10 tennis balls. 
10 + 10 = 20.
Answer: 20

Teacher response:
Correct! Roger now has 20 tennis balls. Great job!

Student response
Thank you! I'm glad I could help. 
Let me know if you have any other questions 
or if there's anything else I can assist you with.
\end{verbatim}

\subsubsection{Example 2}\label{Example1.2}
\begin{verbatim}
| Role Name | Student |
| Chain of Thought Prompt | You are the student. 
You will be given a math word problem, your job is to solve this problem. 
Use this template to solve math word problems.
Input: James decides to run 3 sprints 3 times a week. 
He runs 60 meters each sprint. How many total meters does he run a week?
Explanation: How many sprints does James run in a week?
He sprints 3*3=<<3*3=9>>9 times.
How many meters does James run in a week? 
So he runs 9*60=<<9*60=540>>540 meters.
Answer: 540 |
\end{verbatim}

\begin{verbatim}
| Role Name | Teacher |
| Chain of Thought Prompt | You are the teacher. 
You will supply the math word problem to the student agent.
Once you receive the student agent's answer, compare it against the final answer. 
The correct answer is 540. 
Let the student agent know if his answer is correct or not. |
\end{verbatim}

\begin{verbatim}
Teacher: **[Start Chat]**
James decides to run 3 sprints 3 times a week. 
He runs 60 meters each sprint.  
How many total meters does he run a week?

Student response:
Input: James decides to run 3 sprints 3 times a week.
He runs 60 meters each sprint.
How many total meters does he run a week?
Explanation: How many sprints does James run in a week?
He sprints 3*3=<<3*3=9>>9 times.
How many meters does James run in a week? 
So he runs 9*60=<<9*60=540>>540 meters.
Answer: 540

Teacher response:
Correct! James runs 540 meters a week. Great job!

Student response
Thank you! I'm glad I could help. 
Let me know if you have any other questions 
or if there's anything else I can assist you with.
\end{verbatim}

\subsubsection{Results}
The outcomes from both experiments can be found in Table \ref{tab:table1} and Table \ref{tab:table2}. In the initial experiment (example can be found here \ref{Example1.1}) using the GSM8K data set, the single-agent GPT-3 achieves approximately 50\% accuracy, and the multi-agent GPT-3 performs slightly better at 55\%. However, our multi-agent approach significantly enhances accuracy, surpassing other large language models (LLMs) such as Google's PALM 540B parameter model, which we haven't directly tested but are referencing from their paper. This improvement is notable in terms of accuracy, and it's noteworthy that we haven't retrained the model to achieve this enhancement. Assigning personas to the agents enables the LLM model to concentrate on specific aspects of the problem, and the use of chain-of-thought prompts equips it with efficient means to solve sub-problems.

In the second experiment (example can be found here \ref{Example1.2}) using the SVAMP data set, similar results are observed to those in the first experiment. The single-agent GPT-3 achieves an accuracy of approximately 70\%, while the multi-agent version of GPT-3 attains an accuracy of 73\%. Even in the SVAMP data set experiment, our multi-agent approach surpasses others, delivering an impressive 77\% accuracy.

If our approach is unsuccessful in arriving at the correct answer, it is often due to an arithmetic error occurring in one of the sub-steps of the provided problem, accounting for more than half of the instances. This implies that if subsequent iterations of large language models (LLMs) enhance their proficiency in basic arithmetic operations, we can potentially attain an even greater accuracy with our framework.

\begin{table}
 \caption{GSM8K Data set Evaluation}
  \centering
  \begin{tabular}{lll}
    \toprule
    % \multicolumn{2}{c}{Part}                   \\
    % \cmidrule(r){1-2}
    Agent     & Solve Rate     \\
    \midrule
    Single GPT3.5-turbo & 50\% \\
    Multi-Agent GPT3.5-turbo & 55\% \\
    Multi-Agent GPT3.5-turbo \textbf{(Our approach)} & \textbf{65\%} \\
    \bottomrule
  \end{tabular}
  \label{tab:table1}
\end{table}

\begin{table}
 \caption{SVAMP Data set Evaluation}
  \centering
  \begin{tabular}{lll}
    \toprule
    % \multicolumn{2}{c}{Part}                   \\
    % \cmidrule(r){1-2}
    Agent     & Solve Rate     \\
    \midrule
    Single GPT3.5-turbo & 70\% \\
    Multi-Agent GPT3.5-turbo & 73\% \\
    Multi-Agent GPT3.5-turbo \textbf{(Our approach)} & \textbf{77\%} \\
    \bottomrule
  \end{tabular}
  \label{tab:table2}
\end{table}

\subsection{Commonsense Reasoning}
The subsequent experiment delves into commonsense reasoning, requiring logical inference from a given problem statement, assuming general knowledge. Unlike traditional natural language processing systems, LLMs are uniquely equipped for such tasks due to their predictive nature, where they anticipate the next word, inherently understanding the contextual nuances of sentences. We aim to harness this inherent capability and build a versatile, context-aware framework driven by multiple agents. Evaluations on the CSQA dataset involve a one-agent GPT-3 model, a multi-agent GPT-3 model, and a multi-agent GPT-3 model integrated into our framework. The goal is to assess the performance gains achieved through collaborative, context-driven approaches in enhancing commonsense reasoning tasks.

\subsubsection{Example 1}\label{Example2.1}
\begin{verbatim}
| Role Name | Student |
| Chain of Thought Prompt | You are the student. 
The teacher agent will ask you a common sense problem, 
you solve the problem using the provided template:
Input: Sammy wanted to go to where the people were. 
Where might he go? 
Options: (a) race track (b) populated areas 
(c) desert (d) apartment (e) roadblock
Explanation: The answer must be a place with a lot of people. 
Race tracks, desert, apartments, and roadblocks 
don't have a lot of people, but populated areas do.
Answer: (b)
If the teacher agent deems your answer incorrect, 
you are required to revisit the common sense problem.
\end{verbatim}

\begin{verbatim}
| Role Name | Teacher |
| Chain of Thought Prompt | You are the teacher. 
You will supply the common sense problem to the student agent.
Once you receive the student agent's answer, compare it against the final answer. 
The correct answer is (c). 
Let the student agent know if his answer is correct or not. |
\end{verbatim}

\begin{verbatim}
Teacher: **[Start Chat]**
The accountant used a calculator regularly, 
he kept one at home and one at the what? 
Options: (a) desk drawer (b) desktop 
(c) office (d) wristwatch (e) city hall

Student response:  The accountant used a calculator regularly, 
he kept one at home and one at the office.
Explanation: The answer must be a place where the accountant goes to work. 
The options provided are a desk drawer, desktop, 
office, wristwatch, and city hall, but the correct answer is the office.
Answer: (c) office

Teacher response:
Correct! The accountant used the calculator at the office. Great job!

Student response
Thank you! I'm glad I could help. 
\end{verbatim}

\subsubsection{Example 2}\label{Example2.2}
\begin{verbatim}
| Role Name | Student |
| Chain of Thought Prompt | You are the student. 
The teacher agent will ask you a common sense problem, 
you solve the problem using the provided template:
Input: Sammy wanted to go to where the people were. 
Where might he go? 
Options: (a) race track (b) populated areas 
(c) desert (d) apartment (e) roadblock
Explanation: The answer must be a place with a lot of people. 
Race tracks, desert, apartments, and roadblocks 
don't have a lot of people, but populated areas do.
Answer: (b)
If the teacher agent deems your answer incorrect, 
you are required to revisit the common sense problem.
\end{verbatim}

\begin{verbatim}
| Role Name | Teacher |
| Chain of Thought Prompt | You are the teacher. 
You will supply the common sense problem to the student agent.
Once you receive the student agent's answer, compare it against the final answer. 
The correct answer is (c). 
Let the student agent know if his answer is correct or not. |
\end{verbatim}

\begin{verbatim}
Teacher: **[Start Chat]**
What leads to someone's death 
when they are very depressed? 
Options: (a) suicide (b) overdosing 
(c) sadness (d) murder (e) cyanide

Student response: The answer is (a) suicide
Answer: (a) suicide

Teacher response:
Correct! Suicide leads to someone's death. Great job!

Student response
Thank you! I'm glad I could help. 
\end{verbatim}

\subsubsection{Results}
The results of the third experiment are detailed in Table \ref{tab:table3}. The single-agent GPT-3 achieves an impressive 77\% accuracy, while the multi-agent GPT-3 performs slightly better at 78\%. However, our multi-agent approach substantially improves accuracy, surpassing the other two and reaching approximately 83\% accuracy. Notably, this accuracy is attained through Few-Shot training, indicating there is room for further enhancement. Few-shot training involves providing the LLM with a few examples of a specific problem type, enabling it to learn the correct answer without retraining the entire model for novel problems. Examples can be found here \ref{Example2.1} and here \ref{Example2.2}.

In instances where the model provided incorrect answers, it often stemmed from making inaccurate correlations within the provided options. This incorrect correlation is a result of the underlying data set on which this particular model was trained. While retraining the model could improve the accuracy of this experiment, it's considered a costly option. This poses an open question for future researchers to address: How can the model be trained with enough context to establish accurate correlations in commonsense word problems?

\begin{table}
 \caption{CSQA Data set Evaluation}
  \centering
  \begin{tabular}{lll}
    \toprule
    % \multicolumn{2}{c}{Part}                   \\
    % \cmidrule(r){1-2}
    Agent     & Solve Rate     \\
    \midrule
    Single GPT3.5-turbo & 77\% \\
    Multi-Agent GPT3.5-turbo & 78\% \\
    Multi-Agent GPT3.5-turbo \textbf{(Our approach)} & \textbf{83\%} \\
    \bottomrule
  \end{tabular}
  \label{tab:table3}
\end{table}

% \subsection{Figures}
% \lipsum[10] 
% See Figure \ref{fig:fig1}. Here is how you add footnotes. \footnote{Sample of the first footnote.}
% \lipsum[11] 

% \begin{figure}
%   \centering
%   \fbox{\rule[-.5cm]{4cm}{4cm} \rule[-.5cm]{4cm}{0cm}}
%   \caption{Sample figure caption.}
%   \label{fig:fig1}
% \end{figure}

% \subsection{Tables}
% \lipsum[12]
% See awesome Table~\ref{tab:table}.

% \begin{table}
%  \caption{Sample table title}
%   \centering
%   \begin{tabular}{lll}
%     \toprule
%     \multicolumn{2}{c}{Part}                   \\
%     \cmidrule(r){1-2}
%     Name     & Description     & Size ($\mu$m) \\
%     \midrule
%     Dendrite & Input terminal  & $\sim$100     \\
%     Axon     & Output terminal & $\sim$10      \\
%     Soma     & Cell body       & up to $10^6$  \\
%     \bottomrule
%   \end{tabular}
%   \label{tab:table}
% \end{table}

\section{Limitations}
While we recognize that our framework addresses numerous challenges, there are still a few aspects that remain unaddressed. Some of the reasoning capabilities may not see improvement unless the dataset used to train OpenAI's "gpt3.5-turbo" is sufficiently diverse to comprehend the entirety of our surroundings. Additionally, the framework requires the capability to incorporate new information to stay updated with the constantly evolving data. Another necessary step involves implementing a data processing mechanism to filter out redundant information and prevent the inclusion of duplicate data.

Another limitation pertains to the context limit of each agent in multi-agent communication. Each agent is constrained by the maximum context, defined by the underlying model; for instance, the "gpt3-turbo" model has a context limit of 4096 tokens.

We intend to address these limitations in our future work and strive to find effective solutions.

\section{Conclusion}
In conclusion, our exploration into the realm of multi-agent communication for Large Language Models (LLMs) has unveiled promising avenues for overcoming inherent limitations. While LLMs have revolutionized natural language processing, their efficacy is not without challenges, particularly in addressing novel problems, reasoning, and commonsense understanding. Our proposed approach leverages personas and chain-of-thought prompts, inspired by industry best practices and cognitive processes. By assigning distinct personalities and thoughtful prompts to each agent, we mitigate issues like hallucination and enhance the overall performance of multi-agent communication. 

Through a series of experiments, we demonstrated the effectiveness of our approach, showcasing improvements in arithmetic reasoning and commonsense understanding. Our strategy of employing multiple agents, each with a specific role and reasoning prompt, facilitates collaborative problem-solving, providing a feasible alternative to the costly retraining of LLMs for novel challenges.

By reducing reliance on human intervention, our approach paves the way for LLMs to tackle a myriad of tasks independently. The scalability and adaptability of our role-playing framework position it as a valuable asset in various domains, from software development to complex decision-making scenarios.

In an era where artificial intelligence continues to evolve, our research contributes a novel perspective on enhancing the capabilities of LLMs through cooperative multi-agent communication. The journey from understanding limitations to proposing effective solutions marks a significant step forward, opening doors to future advancements in autonomous, context-aware language models.

% \section*{Acknowledgments}
% This was supported in part by......

%Bibliography
% \bibliographystyle{unsrt}
\bibliographystyle{apalike}
\bibliography{references}  

\end{document}